\documentclass[aps,prl,twocolumn,showpacs]{revtex4}
\usepackage{graphicx}
\newcommand{\ds}{ _{\downarrow}}
\newcommand{\us}{ _{\uparrow}}
\newcommand{\up}{\uparrow}
\newcommand{\down}{\downarrow}
\newcommand{\gzu}{ G_{0 \uparrow}}
\begin{document}
\draft \title{Normal state of highly polarized Fermi gases: simple
many-body approaches}
\author{R. Combescot} \address{Laboratoire de
Physique Statistique, Ecole Normale Sup\'erieure, 24 rue Lhomond,
75231 Paris Cedex 05, France}
\author{A. Recati and C. Lobo}
\address{Dipartimento di Fisica, Universit\`a di Trento and CNR-INFM
BEC Center, I-38050 Povo, Trento, Italy}
\author{F. Chevy}
\address{Laboratoire Kastler Brossel, Ecole Normale Sup\'erieure, 24
rue Lhomond, 75231 Paris Cedex 05, France}
\date{Received \today}

\begin{abstract}
We consider the problem of a single $\down$ atom in the presence
of a Fermi sea of $\up$ atoms, in the vicinity of a Feshbach
resonance. We calculate the chemical potential and the effective
mass of the $\down$ atom using two simple approaches: a many-body
variational wave function and a $T$-matrix approximation. These
two methods lead to the same results and are in good agreement
with existing quantum Monte-Carlo calculations performed at
unitarity and, in one dimension, with the known exact solution.
Surprisingly, our results suggest that, even at unitarity, the
effect of interactions is fairly weak and can be accurately
described using single particle-hole excitations. We also consider
the case of unequal masses.
\end{abstract}

\maketitle

The investigation of ultracold Fermi gases with two unbalanced
hyperfine states (which we shall denote as $\up$ and $\down$) has
been through an impressive expansion last year. This subfield is of
great interest both on practical and on theoretical grounds. Indeed on
one hand it is related to other fields of physics, namely
superconductivity, astrophysics and high-energy physics where similar
situations arise \cite{cana}. On the other hand the additional
parameter provided by the population imbalance should provide a tool
to deepen our understanding of the BEC-BCS crossover in these systems
and contribute to improve our control of many-body theory. This recent
activity has been started by striking experimental results
\cite{rimit} which have given rise to a considerable number of
theoretical works \cite{shra}. Experiments performed on trapped
systems have observed equal density superfluid states as well as
partially and fully polarized regions.

The analysis of the $T=0$ phase separation requires the knowledge of
the properties of both the superfluid and the partially polarized
normal phase \cite{bulgac,fred,lrgs}. This
has been developed in the recent work of Lobo \emph{et al} \cite{lrgs}
where, at unitarity, the properties of the partially polarized
phase have been obtained by calculating through a quantum Monte-Carlo
(MC) approach the parameters which characterize a single $\down$
atom immersed in a Fermi sea of $\up$ atoms, with density
$n\us=k_F^{3}/(6\pi ^2)$. In this way they were able to obtained a
very good agreement with experimental results.

Here we consider the general problem of a single $\down$ atom in a
completely polarized $\up$ atom Fermi sea. The $\up - \down$
interaction is characterized by an s-wave scattering length $a$ whose
value can be tuned via a Feshbach resonance from the BEC ($1/k_Fa \gg
1$) to the BCS ($1/k_Fa \ll -1$) limits. The Fermi gas is assumed to
be ideal due to the suppression of higher angular momentum scattering
at low temperatures. This problem is a much simpler one than the case of
two equal spin populations in the BEC-BCS crossover, although still
quite nontrivial due to the absence of a small parameter in the
strongly interacting regime.  It is the simplest realization of the
moving impurity problem and it bears a strong similarity with other
old, famous and notoriously difficult condensed matter problems, such
as the Kondo problem, the X-ray singularity in metals \cite{mahan} and
the mobility of ions \cite{prokofev} and $^4$He atoms \cite{boronat}
in $^3$He. This gives to the present atomic system a much wider
significance beyond the context of polarized Fermi gases. We may hope
to have a full control and understanding of this system, and to obtain
further physical insight in these kind of problems.

In this paper, in addition to considering a general scattering length
$a$, we will extend the parameter range by treating the general case
where the masses $m\us$ and $m\ds$ are different, although they are
the same in experiments up to now. This can be accomplished by using
atoms belonging to different elements. The limit $m_\down/m_\up
\rightarrow \infty$ is then compared to the fully solvable problem of
a fixed impurity in a Fermi sea.

Here we are interested in the two physical quantities which have been
calculated by Lobo \emph{et al} \cite{lrgs}, namely the chemical
potential $\mu \ds$ of the $\down$ atom and its effective mass
$m^{*}$. We have addressed this problem by two different and
complementary many-body methods. The first one is a natural extension
of the many-body trial wave function used by Chevy \cite{fred} to
obtain the effective mass. The second one is a $T$-matrix
approximation, whose basic ingredient, ladder diagrams, appear in more
elaborate schemes. This approximation has already been used in a wide
range of physical systems, including high $T_c$ superconductors
\cite{max} and the BEC-BCS crossover \cite{piestr,clk}, and it is
known to give reasonable results. The variational method has the
advantage of providing a rigorous upper bound to the energy while the
$T$-matrix can be easily extended to include more sophisticated
approximations. As we will see these two methods lead exactly to the
same results and moreover, are in surprisingly good agreement with MC
calculations \cite{lrgs,pilati}.

The trial wave function $|\psi\rangle$ we consider, for a system of
total momentum ${\bf p}$, is the following momentum eigenstate (we set
$\hbar=1$ throughout the paper):
\begin{eqnarray}\label{var}
|\psi\rangle= \phi_0 |{\bf p}\rangle\ds\;|0\rangle\us + \!
\sum_{q<k_F}^{k>k_F}\phi_{{\bf q}{\bf k}} |{\bf p}+{\bf q}-{\bf
  k}\rangle\ds\;c^{\dag}_{{\bf k}\us}\,c_{{\bf q}\us}\,|0\rangle\us
\end{eqnarray}
where $c_{{\bf k}\us}$ and $c^{\dag}_{{\bf k}\us}$ are
annihilation and creation operators. In the first term the
$\up$-spin free Fermi sea is in its ground state $|0\rangle\us=
\prod_{k<k_F} c^{\dag}_{{\bf k}\us}\,|vac\rangle$ and the
$\down$-spin atom is in the plane-wave state $|{\bf p}\rangle\ds$,
while in the second term it is in excited states corresponding to the
creation of a particle-hole pair in the Fermi sea with momentum
${\bf k}$ and ${\bf q}$ respectively, the $\down$-spin atom
carrying the rest of the momentum. The coefficients $\phi_0$ and
$\phi_{{\bf q}{\bf k}}$ are found by minimizing the total energy.
This wave function is by construction suitable to reproduces the 
molecule in the BEC limit as well as the perturbative mean-field limit. 
We follow the procedure of Ref.\cite{fred}, in
particular with respect to the handling of the zero range
interaction potential and the corresponding regularization in
terms of the scattering length. We obtain for the change in energy
$E$ due to the addition of the $\down$-spin atom:
\begin{eqnarray}\label{vareq}
\hspace{-17mm}E=\epsilon _{\downarrow p}+ \sum_{q<k_F}f(E ,{\bf
p},{\bf q})
\end{eqnarray}
\vspace{-5mm}
\begin{eqnarray}\label{def}
\frac{1}{f(E ,{\bf p},{\bf q})}\!=\!\frac{m_{r}}{2\pi a}-
\!\sum_{k}\!\frac{2m_{r}}{k^{2}}\!+\!\!\!\sum_{k>k_F} \frac{1}
{\epsilon_{\uparrow k}\!+\!\epsilon _{\downarrow{\bf p}\!+\!{\bf
q}\!-\!{\bf k}}\!-\!\epsilon_{\uparrow q}\!-\!E} \nonumber
\end{eqnarray}
where $\epsilon_{\uparrow,\downarrow k} =
k^{2}/2m_{\uparrow,\downarrow}$ is the kinetic energy of the $\up$ and
$\down$ atoms, and $m_r = m\us m\ds /(m\us + m\ds)$ the reduced
mass. For ${\bf p}=0$ we have $E=\mu \ds$, while the variation of $E$
for small ${\bf p}$ gives the effective mass.

These results can be obtained exactly from the knowledge of the
self-energy $\Sigma (p,\omega)$ of the $\down$ atom. Indeed the
pole of the $\down$ Green's function $G\ds(p,\omega )=[\,\omega
-\epsilon _{\downarrow p} + \mu \ds - \Sigma (p,\omega)]^{-1}$,
giving the dispersion relation of the $\down$ quasiparticle,
satisfies:
\begin{eqnarray}\label{disprel}
\omega -\epsilon_{\downarrow p} + \mu \ds -  \Sigma (p,\omega) = 0
\end{eqnarray}
 Since $\omega $ is the energy measured from the chemical potential,
 and that physically the chemical potential corresponds to the
 addition of a particle with zero momentum $p=0$, the chemical
 potential is given by:
\begin{eqnarray}\label{chem}
\mu \ds =  \Sigma (0,0)
\end{eqnarray}
Moreover the effective mass, giving the dispersion relation of the
$\down$ quasiparticle, is obtained from the small $p$ behaviour of the
dispersion relation and is given by:
\begin{eqnarray}\label{mstar}
\frac{m^*}{m\ds} = \frac{1-\frac{\partial \Sigma}{\partial \omega
}}{1+2m\ds \frac{\partial \Sigma}{\partial p^{2}}}
\end{eqnarray}
where the derivatives of the self-energy are taken for $p=\omega =0$
($\Sigma$ is real in the situations we deal with below).

The self-energy itself is obtained \cite{agd} from the unknown
two-particle vertex $\Gamma$, the only very important simplifying
feature of our problem being that the $\up$-atom Green's function is
exactly the bare one, namely $\gzu (k,\omega )= [\omega -\epsilon
_{\uparrow k} + \mu \us]^{-1}$, where $\mu \us=k_F^{2}/2m\us$, since
the single $\down$ atom does not perturb in the thermodynamic limit
the free Fermi sea of $\up$ atoms. To obtain actual answers, we
proceed to take the simplest approximation for the two-particle
vertex, namely the $T-$matrix approximation. In this approximation the
$\down$ atom interacts only with a single $\up$ atom. This implies in
particular that the only excited states of the $\up$ Fermi sea coming
in this problem are the single particle-hole excitations, just as in
Eq.(\ref{var}). In this $T-$matrix approximation the $\down$ and the
$\up$ atoms scatter any number of times through the bare potential,
just as in any two-body problem solved exactly by perturbation theory,
which leads to the well-known series of "ladder" diagrams. When there
is a single $\up$ atom the problem reduces to the full scattering of
the $\down$ and $\up$ atoms, and the solution can be expressed in
terms of the scattering amplitude. When this problem is compared to
the one in the presence of the Fermi sea of $\up$ atoms, the bare
interaction $V$ can be eliminated in favor of the scattering
properties. In our approximation the vertex $\Gamma$ depends only on
the total momentum ${\bf K}$ and the total energy $\Omega $ of
incoming particles. After performing the above steps we find
\cite{clk,agd}:
\begin{eqnarray}\label{eqgam1}
\Gamma^{-1} (K,\Omega ) = \frac{m_r}{2\pi a}- \int \frac{d{\bf k}}{(2\pi )^{3}}
 \;\left[\,\frac{2m_r}{k^2}\right. +\hspace{20mm}\\ \nonumber
\hspace{3mm} \left. G\ds \left(k,\Omega +\mu \us - \epsilon_{\uparrow
{\bf K}-{\bf k}}\right)\:\theta\left(\epsilon_{\uparrow {\bf K}-{\bf
k}}-\mu \us\right) \right]
\end{eqnarray}
where $\theta(x)$ is the Heaviside function. In obtaining this result
we have made explicit use of the fact that we will find $\mu \ds <0$
as it is obvious physically.

In our approximation the self-energy is given by:
\begin{eqnarray}\label{selflad}
\Sigma (p,\omega)\! =\!\frac{1}{2i\pi } \! \int \!\frac{d{\bf
K}}{(2\pi )^{3}} \!\int_{C}\! \!d\Omega \,\Gamma (K,\Omega )\,\gzu
({\bf K}\!-\!{\bf p},\Omega \!-\!\omega )
\end{eqnarray}
where, as mentioned above, the fact of using $\gzu$ as the $\up$-atom
Green's function does not imply any approximation. Contour $C$ goes
anticlockwise around the left-hand side part ${\rm Re}\,\omega <0$ of
the $\omega $ complex plane. We deform it to enclose only the
singularities of the integrant. Physically the singularities of
$\Gamma (K,\Omega )$ correspond to the continuous spectrum of the
scattering states of the $\up$ and $\down$ atom, and possibly to a
bound state of these atoms. On the other hand $\gzu(k,\omega )$ has
just, for fixed $k$, a pole at $\omega =\epsilon_{\uparrow k} - \mu
\us$. Since we work at $T=0$, the particle-particle continuous
spectrum does not contribute as it is physically obvious. Indeed, from
Eq.(\ref{eqgam1}), the corresponding singularities are at $\Omega > 0$
(because of the Heaviside function) and are outside contour $C$. On
the other hand we will mainly restrict ourselves to the case where
there is no bound states between the $\up$ and the $\down$ atoms. Such
a bound state exists clearly in the BEC limit $1/k_Fa\rightarrow
+\infty$ where molecules will be present. While as we shall shortly
show our simple approach recovers also this limiting behaviour, the
effect of a bound state in the intermediate regime will be
investigated in further work.  This leaves us only with the
contribution of the pole of $\gzu$:
\begin{eqnarray}\label{sigm}
\Sigma (p,\omega)\! =\!\!\int\! \!\frac{d{\bf K}}{(2\pi )^{3}}
\,\theta\!\left(\mu \us-\epsilon_{\uparrow {\bf K}\!-\!{\bf
p}}\right)\Gamma (K,\omega +\epsilon_{\uparrow {\bf K}-{\bf
p}}\!-\!\mu \us)\, \\ \nonumber = \frac{1}{2\pi ^2} \int_{0}^{k_F}dK
\;K^{2}\,\langle\Gamma ({\bf K}+{\bf p},\omega +\epsilon_{\uparrow K}-\mu
\us)\rangle
\end{eqnarray}
where the bracket is for the angular average over the direction of ${\bf K}$.

At this stage it is interesting to consider the weak coupling limit $a
\rightarrow 0_{-}$, in which case the first term dominates in the
right-hand side of Eq.(\ref{eqgam1}). This gives $\Gamma (K,\Omega ) =
2\pi a/m_r$ and $\Sigma (p,\omega)=\Sigma (0,0)=\mu \ds=k_F^{3}a/(3\pi
m_r )= 2\pi n\us a/m_r$. Hence we recover the expected result for the
mean-field interaction energy with a short-range interaction $V({\bf
r})= (2\pi a/m_r)\, \delta({\bf r})$. We can view the more general
result Eq.(\ref{sigm}) as having a similar physical interpretation,
but with the effective interaction having now a wavevector and energy
dependence.

In the general case Eq.(\ref{sigm}) together with Eq.(\ref{eqgam1})
provides an integral equation for $\Sigma (p,\omega)$. We leave the
exact solution of this equation for further work, whereas in the
present paper we will proceed to a further approximation. We will just
stop at the first step of an iteration loop which would provide the
full answer, that is we will set $\Sigma (p,\omega)=0$ in the
expression of $G\ds(p,\omega )$. In this case we can check that
Eq.(\ref{vareq}) and (\ref{disprel}) are identical, provided $E$ is
identified with $\omega + \mu \ds$.

For the case of the $\down$ atom chemical potential, we are just
left with solving numerically an equation for $\mu \ds$. At
unitarity $1/k_Fa=0$ and for $m_\down=m_\up$ this gives $\mu
\ds=-0.6066\, \mu \us$, in remarkable agreement with the QMC
\cite{lrgs,fred} result $\mu \ds=-(3/5) (0.97\pm0.02)\, \mu
\us=-0.58\pm0.01\, \mu \us$. This surprising agreement suggests
that the effect of interactions is weak even at unitarity.

It is interesting to investigate the regime where the ratio
$\rho=|\mu \ds|/\mu \us$ becomes large. In this case the
expression of $\Gamma(K,\Omega )$ becomes quite simple. We find to
dominant order $\Gamma(K,\Omega )=[2\pi /(m_r k_F )] [1/k_F a
-(\rho r/(1+r))^{1/2}]^{-1}$, with $r=m\ds /m\us$, leading to the
following explicit equation for the relation between $\rho$ and
$1/k_F a$:
\begin{eqnarray}\label{massapprox}
\frac{1}{k_F a}=\sqrt{\frac{\rho r}{1+r}}-\frac{2}{3\pi }
\frac{1+r}{\rho r}
\end{eqnarray}
\begin{figure}
\vspace{-2mm}
\hspace{-5mm}
\scalebox{0.9}{\includegraphics[width=10cm]{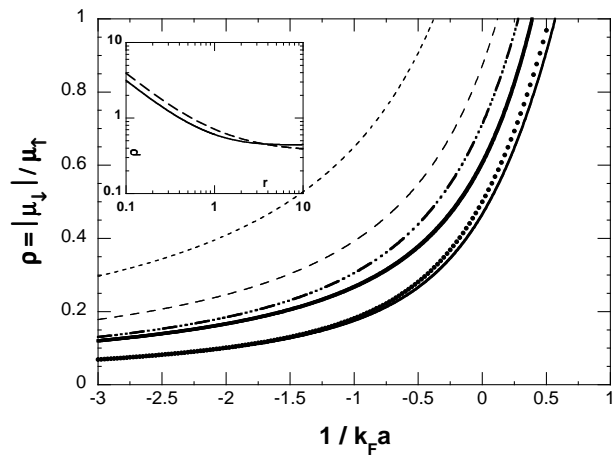}}
\caption{Reduced $\down$ atom chemical potential $|\mu \ds|$ as a
function of $1/k_F a$ for various mass ratios $r=m\ds /m\us$. From top
to bottom $r$ = 0.25 , 0.5, 1. (full thick line, the interpolating
approximation Eq.(\ref{massapprox}) being the thick dashed-triple
dotted line just above), and $\infty$ (lower full thick line), the
exact result Eq.(\ref{infmassexact}) being the dotted curve just
above. The inset compares, at unitarity, the approximation
Eq.(\ref{massapprox}) (dashed line) with the numerical results (full
line) as a function of the mass ratio $r$.}
\label{fig1}
\vspace{-6mm}
\end{figure}
In the case of equal masses $m\us=m\ds$, this is plotted for
comparison in Fig.\ref{fig1} and is seen to give a quite good
agreement with the numerical value \cite{note}. In the weak
coupling regime $a\rightarrow 0_{-}$ we recover the mean field result
given above, while the asymptotic behaviour for large $\rho$ is
$\rho=(1+1/r)(1/k_F a)^{2}$, which is the two-body bound
state energy. This formula can be seen as an interpolation between
these two extremes. At unitarity it gives $\rho=(1+1/r)(2/3\pi
)^{2/3}$. For equal masses we get $\rho \approx 0.71$ which is
fairly near the numerical result.

In the inset of Fig.\ref{fig1} we present the results of our model at
unitarity as a function of the mass ratio $r=m \ds /m \us$. Again the
interpolation Eq.(\ref{massapprox}), i.e. $\rho = (2/3\pi )^{2/3}
(1+1/r)$, is in quite reasonable agreement with numerical results. For
small $r$ the chemical potential $\mu \ds$ goes to $-\infty$, as it
can be seen easily from Eq.(\ref{sigm}) \cite{note2}. In the other limit $m \ds
\rightarrow \infty$ this ratio is seen to saturate.

Assuming that, for the purpose of calculating the chemical potential,
the thermodynamic and infinite mass limits commute, the problem
reduces to that of an impurity interacting with a free Fermi sea,
which is well-known in solid state physics \cite{fried}. It can be
solved exactly in the following way. The Fermi sea is enclosed in
a large sphere of radius $R$, with $R \rightarrow \infty$ in the
thermodynamic limit and the impurity at the center. Since the s-wave
functions have to be zero at the sphere, the allowed wavevectors $k_p$
are given by $k_pR+\delta_0(k_p)=p\pi $ with integer $p \le n$ and
$k_F R=n \pi $. For low energy atoms the phase shift is given by $\tan
\delta_0(k)=-ka$. The energy of all the atoms of the Fermi sea is $E=
\sum_{p} k_p^{2}/2m\us$. The calculation is conveniently performed by
finding the change in energy due to a change in the scattering
length. In this limiting case the change in total energy of the Fermi
sea is identified with the chemical potential of the $\down$ atom. We
find in this way:
\begin{eqnarray}\label{infmassexact}
\hspace{-2mm}\rho \!\equiv \!\frac{|\mu \ds|}{\mu \us} \!= \!\frac{1}{\pi
}\!\left[(1+y^2)(\frac{\pi }{2}\!+\!\arctan y)\!+\!y \right]
\hspace{3mm} y\!=\! \frac{1}{k_F a}
\end{eqnarray}
The result is plotted in Fig.\ref{fig1} and is seen to be in excellent
agreement with the variational and $T$-matrix results. In particular
we find $\rho=0.5$ at unitarity, to be compared to the
approximate value $0.465$.
\begin{figure}
\hspace{-5mm}
\scalebox{0.9}{\includegraphics[width=10cm]{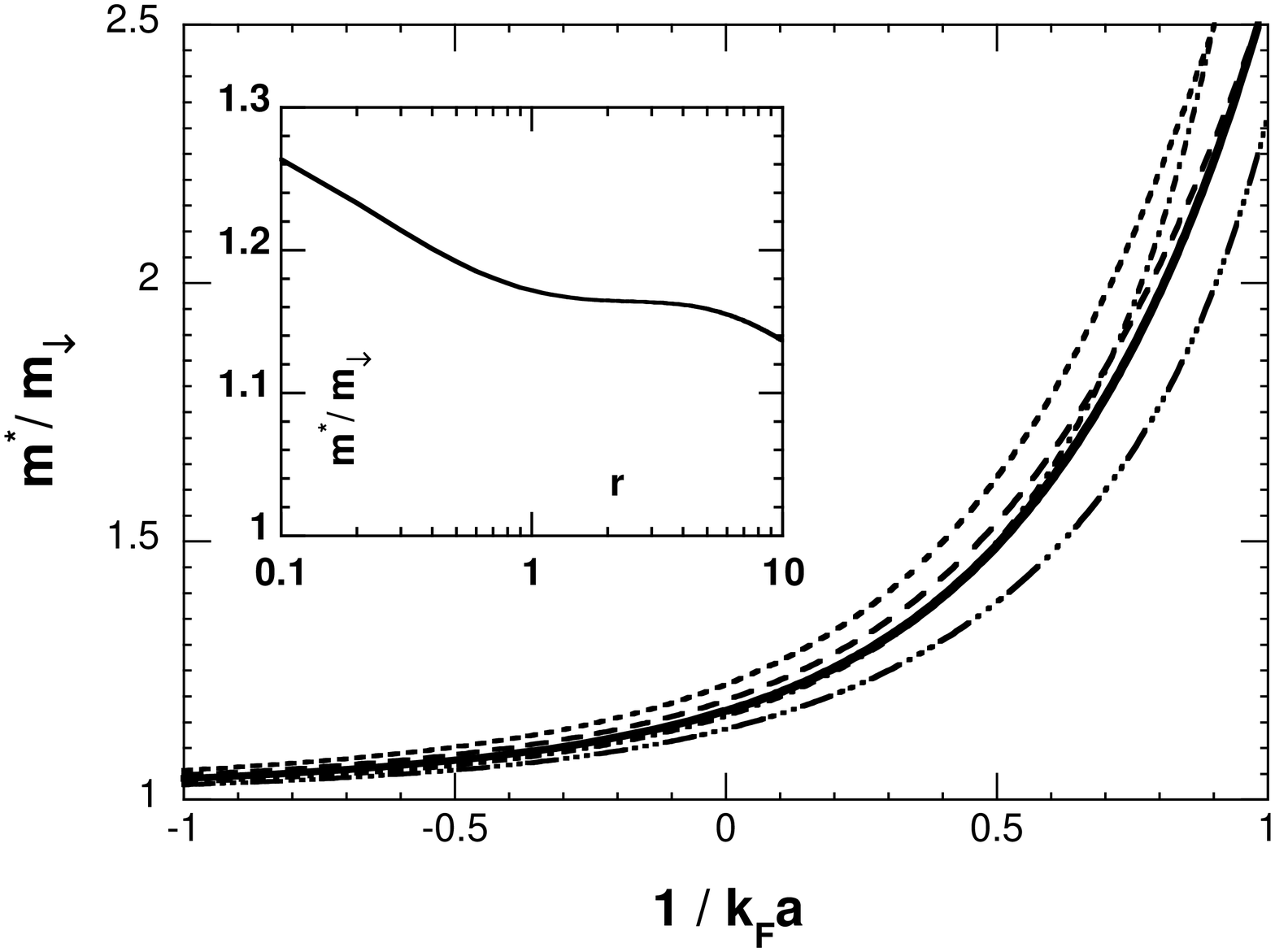}}
\caption{Relative effective mass $m^{*} /m\ds$ as a function of $1/k_F
a$ for various mass ratios $r=m\ds /m\us$. Same conventions as in
Fig.\ref{fig1} for $r$=0.25 , 0.5, 1. The dashed-dotted is $r$=4. and
the dashed-triple dotted is $r$=10. The inset shows $m^{*} /m\ds$ as a
function of $r$ at unitarity.}
\label{fig2}
\vspace{-5mm}
\end{figure}

Finally the relative effective mass $m^{*}/m\ds$ is displayed in
Fig.\ref{fig2} for various mass ratios. The first striking feature
is that the mass enhancement is quite small around unitarity,
whereas we might have expected a much stronger effect around
resonance. Naturally when we go further to the BEC side
$m^{*}/m\ds$ increases rapidly. Quantitatively our result
$m^{*}/m\ds=1.17$ for equal masses at unitarity is in quite
reasonable agreement with the QMC result $1.04(3)$, taking into
account that effective mass is more sensitive to approximations
than energy. The other noticeable feature of Fig.\ref{fig2} is the
weak dependence of $m^{*}/m\ds$ on the mass ratio $r$, as can be
seen in the inset at unitarity. We note that, within our
approximation, no bound state appears in the plotted range. We
have to keep in mind that, in improved approximations, the
location for the appearance of a bound state might be somewhat
changed.

To further check the reliability of our approach we have done the
calculations in one dimension. In this case we can compare our
results with the exact solution for equal masses \cite{McGuire}.
We find a very good agreement along the whole crossover, from
BCS-like state to the molecular Tonks state, for the energy, while
the mass is more sensitive to our approximations precisely when
the two-body bound state plays a major role.

In conclusion we have found that the physical properties of a
single $\down$ atom in the presence of a Fermi sea of $\up$ atoms
can be described fairly accurately by the simple inclusion of
single particle-hole excitations. Two equivalent schemes based
respectively on a many-body trial wave function and simple
$T$-matrix approach were developed and give very good agreement
with known MC results for $m\us=m\ds$ at unitarity.

We are grateful to S. Giorgini, M. Yu. Kagan, X. Leyronas and S.
Stringari for stimulating discussions. The ``Laboratoire de
Physique Statistique'' and ``Laboratoire Kastler Brossel'' are
``Laboratoires associ\'es au Centre National de la Recherche
Scientifique et aux Universit\'es Paris 6 et Paris 7''. C. L. and
A. R. acknowledge support by the Ministero dell'Istruzione,
dell'Universit\`a e della Ricerca (MIUR). F. C. acknowledges
support from Region Ile de France (IFRAF).


\begin{references}
\bibitem{note} Ref. \cite{pilati} has calculated $\rho$ on the $a<0$
  side for equal masses and their results are in excellent agreement
  with ours. The agreement is not as good for the effective mass, but
  still within 10 per cent in the worst case.
\bibitem{note2} It is known that, for large mass imbalance, few-body
  phenomena might play a significant role in the many-body physics of
  a mixture. However here we will restrict ourselves to our ansatz and
  not take this type of correlations into account.
\end{references}
\end{document}